\documentclass[showpacs,preprintnumbers,amsmath,amssymb,prl,aps,twocolumn]{revtex4}

\usepackage{graphicx}
\usepackage{dcolumn}
\usepackage{bm}
\usepackage[usenames,dvipsnames]{color}
\usepackage{ulem}

\newcommand{\lico}{Li$_2$CuO$_2$}
\newcommand{\cugeo}{CuGeO$_3$}
\newcommand{\cuo}{CuO$_4$}

\newcommand{\CGO}{CuGeO$_3$}
\newcommand{\LCO}{Li$_2$CuO$_2$}

\begin{document}
\bibliographystyle{prsty}

\title{Determining the Short-Range Spin Correlations in Cuprate Chain Materials with Resonant Inelastic X-ray Scattering}

\author{Claude Monney$^{1}$}
\author{Valentina Bisogni$^2$}
\author{Ke Jin Zhou$^1$}
\author{Roberto Kraus$^2$}
\author{Vladimir N. Strocov$^1$}
\author{G\"unter Behr$^2$\footnote{Deceased}}
\author{Ji{\v r}i M\'alek$^{2,3}$}
\author{Roman Kuzian$^{2,4}$}
\author{Stefan-Ludwig Drechsler$^2$}
\author{Steve Johnston$^{2}$}
\author{Alexandre Revcolevschi$^5$}
\author{Bernd B\"uchner$^{2,6}$}
\author{Henrik M. R\o nnow$^7$}
\author{Jeroen van den Brink$^{2,6}$}
\author{Jochen Geck$^2$}
\email{j.geck@ifw-dresden.de}
\author{Thorsten Schmitt$^1$}
\email{thorsten.schmitt@psi.ch}

\affiliation{%
$^1$Research Department Synchrotron Radiation and Nanotechnology, Paul Scherrer Institut, CH-5232 Villigen PSI, Switzerland\\
$^2$Leibniz Institute for Solid State and Materials Research IFW-Dresden, Helmholtzstrasse 20, D-01171 Dresden, Germany\\
$^3$Institute of Physics, ASCR, Na Slovance 2, CZ-18221 Praha 8, Czech Republic\\
$^4$Donostia International Physics Center (DIPC), Donostia-San Sebastian, Spain\\
$^5$Laboratoire de Physico-Chimie de l'Etat Solide, ICMMO, Universit\'e Paris-Sud, 91405 Orsay Cedex, France\\
$^6$ Department of Physics, TU-Dresden, D-01062 Dresden, Germany\\
$^7$Laboratory for Quantum Magnetism, ICMP, Ecole Polytechnique F\'ed\'erale de Lausanne (EPFL), Switzerland
}%

\date{\today}

\begin{abstract}
We report a high-resolution resonant inelastic soft x-ray scattering study 
of the quantum magnetic spin-chain materials \lico\ and \cugeo. By tuning 
the incoming photon energy to the oxygen $K$-edge, a strong excitation around 
3.5\,eV energy loss is clearly resolved for both materials. Comparing the 
experimental data to many-body calculations, we identify this excitation as 
a Zhang-Rice singlet exciton on neighboring \cuo-plaquettes. We 
demonstrate that the strong temperature dependence of the inelastic
scattering related to this high-energy exciton enables to probe short-range spin correlations on the 1\,meV scale 
with outstanding sensitivity.
\end{abstract}

\maketitle

Two-dimensional cuprate materials play an essential role in condensed matter physics as they show high temperature superconductivity upon charge doping. For better understanding these complex materials, it is important to tackle simpler model systems sharing similar key components and showing reduced complexity, namely one dimensional cuprate chains made out of CuO$_4$ plaquettes.
In this context, Zhang-Rice singlets (ZRS) 
are fundamental elementary excitations being composite
objects generic to hole doped or photon excited strongly correlated
charge transfer insulators, and that are especially well-known 
in the cuprates \cite{Zhang88}.
However, more than two decades after their theoretical
discovery and numerous observations afterwards, 
there is still significant theoretical and experimental activity aimed at clarifying their complex
details e.g. with respect to additional orbitals \cite{Chen10}
or specific magnetic correlations beyond their  
centers \cite{Morinari12}. 

\begin{figure}
\centering
\includegraphics[width=9cm]{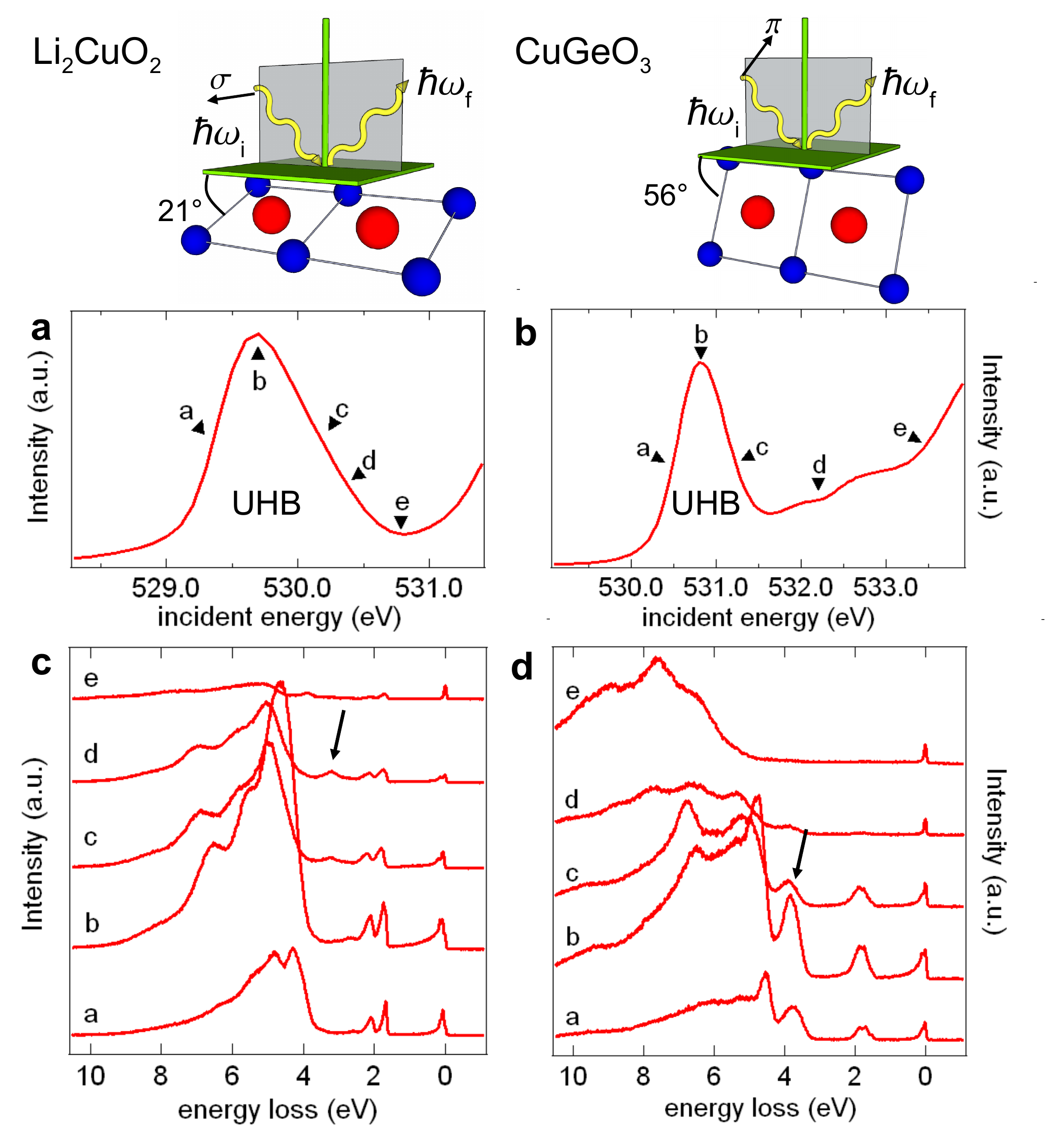}
\caption{\label{fig_1} (a) XAS measured at the oxygen $K$-edge for
 \lico\ with $\sigma-$polarized light at 20K and (b) for \cugeo\ 
 measured with $\pi-$polarized light at 40K (using the total fluorescence 
 yield). Pre-edge peaks are related to the upper Hubbard band. (c) RIXS 
 spectra (on an energy loss scale) measured at the oxygen $K$-edge 
 for \lico\ with $\sigma-$polarization at 20K and (d) for \cugeo\ with 
 $\pi-$polarization at 40K. The incident energies used for the different 
 spectra are indicated by full triangles on the corresponding XAS spectra 
 in graphs a and b.  The RIXS spectra are normalised to the acquisition time.
}
\end{figure}

Edge-shared cuprate chains represent a particular class of quantum magnets in which the local 
 geometry gives rise to competing nearest ferromagnetic
 (FM) or antiferromagnetic (AFM) exchange coupling $J_1$ 
and  frustrating next-nearest neighbor AFM $J_2$ superexchange couplings. 
The AFM one-dimensional spin-1/2 $J_1$-$J_2$ Heisenberg model describes such frustrated magnetic interactions, due to which quantum fluctuations can 
alter both ground state and spin correlations\,\cite{Mikeska1Dmagn}. 
Generalizing this model by varying the signs and ratio 
of $J_1$ and $J_2$ gives rise to a rich phase diagram with ground states 
spanning FM, 
AFM, helical, and gapped singlet states. In real 
materials, the presence 
of interchain coupling and occasionally coupling to the lattice adds 
complexity to this behavior, rendering theoretical treatment more 
difficult\,\cite{Drechsler2009}. It is therefore desirable to obtain 
experimental access to nearest neighbor spin correlations both within 
the ground state probed at very low-temperature
and in thermally occupied excited spin-states
as a function of temperature ($T$) 
\cite{LMOdd}.
Both \lico\ and \cugeo\ realize frustrated edge-shared chain 
systems that exhibit ground states with completely different intrachain spin 
correlations. While \cugeo, on the one hand, displays the well-established 
spin-Peierls phase below $T_{SP}=14$\,K resulting in a gapped singlet state 
with pronounced AFM nearest neighbor spin correlations in the chain direction 
\cite{HaseSP,CastillaQM}, \lico, on the other hand, shows FM
long-range spin-order along the chains below $T_{N}=9$\,K \,\cite{LorenzINS}.

In this letter, we demonstrate that resonant inelastic x-ray scattering (RIXS) at the oxygen $K$-edge allows to probe ZRS
excitations \cite{MalekLCO,MatiksLCV} for these two quantum magnetic spin-chain materials, \lico\ and \cugeo\ with unique sensitivity. Comparing the experimental results to theoretical calculations, we also show that these excitations display an extraordinarily strong temperature dependence, which is directly
related to the spin texture of the studied materials.
This effect together with the high sensitivity of RIXS is shown to be 
a powerful probe to study nearest and next nearest neighbor spin correlations in cuprate chains.

RIXS experiments were performed at the ADRESS beamline \cite{beamline} of 
the Swiss Light Source, Paul Scherrer Institut, using the SAXES spectrometer 
\cite{SAXES}. RIXS spectra were recorded in typically 2h acquisition time, 
achieving a statistics of 100-150 photons on the peaks of interest. A 
scattering angle of $130^\circ$ was used and all the spectra were measured 
at the specular position, i.e. at an incidence angle of $65^\circ$ (see e.g. Fig. 1 in Ref. \cite{Braicovich2010} for a sketch of the scattering geometry), meaning 
that no light momentum is transferred to the system along the chain 
direction. The combined energy resolution was 60 meV at the oxygen 
$K$-edge ($\sim530$ eV). \lico\ single crystals (which are hygroscopic 
crystals) \cite{BehrLCO} were cleaved in-situ at the pressure of about 
$5\cdot10^{-10}$ mbar and at 20\,K, while \cugeo\ single crystals were 
cleaved at $10^{-7}$ mbar and RT, producing both mirror like surfaces. 
In the case of \lico, the surface is perpendicular to the [101] axis, so 
that the \cuo\ plaquettes are 21$^\circ$ tilted away from the surface. 
In \cugeo\ \cite{RevcoCGO}, the surface is oriented perpendicular to the 
[100] axis, so that the \cuo\ plaquettes are 56$^\circ$ tilted away from 
the surface. 

RIXS probes low-energy charge, spin, orbital and lattice excitations of 
solids \cite{AmentReview,SchlappaNature}. The RIXS process is based on 
the coherent absorption and reemission of photons. The incoming photon 
with energy  $\hbar\omega_i$ virtually excites the electronic system from 
an initial state $|i\rangle$ to an intermediate state $|m\rangle$, which 
then decays again into a final state $|f\rangle$ by emitting an outgoing 
photon with energy $\hbar\omega_f$\,\cite{AmentReview}. 
We tuned $\hbar\omega_i$ to the oxygen $K$ pre-edge, as shown by full 
triangles on the x-ray absorption spectra (XAS) in Figs.\,\ref{fig_1}\,a 
and b. At this energy, O\,$1s$ core electrons are directly excited into 
the so-called upper Hubbard band (UHB) \cite{NeudertXAS,CorradiniXAS}, 
which yields a strong resonant enhancement of electronic excitations 
involving hybridised Cu\,$3d$ and O\,$2p$ valence states. Choosing 
different incident energies corresponds to exciting different intermediate 
states in the RIXS process.

\begin{figure}
\centering
\includegraphics[width=8.8cm]{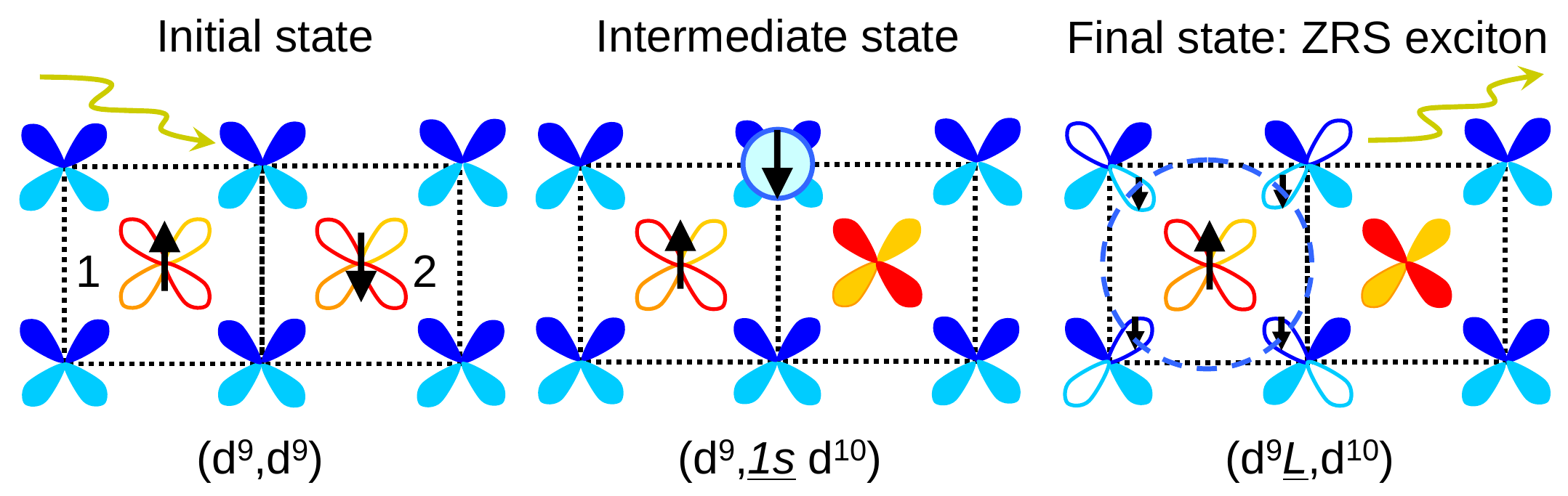}
\caption{\label{fig_2} Schematic illustration of how a ZRS exciton is 
created in the RIXS process at the oxygen $K$-edge. See text for 
detailed explanations. Unoccupied states are depicted with empty orbitals.}
\end{figure}

In  Figs.\,\ref{fig_1}\,c and d, RIXS intensities for \lico\ and \cugeo\ 
are plotted as a function of the photon-energy loss 
$\hbar\Omega=\hbar\,(\omega_i-\omega_f)$. The unprecedented energy 
resolution of these data reveals remarkably rich spectra, exhibiting 
different sharp peaks. For both materials intense and broad structures 
are observed at $\hbar\Omega>4.5$\,eV that shift with $\hbar\omega_i$ 
and can be identified as conventional x-ray fluorescence \cite{KotaniReview}. 
These transitions will not be considered in the following.  
Instead we will focus on the excitations observed at $\hbar\Omega<4.5$\,eV. 
These excitations occur at fixed energy losses and have the largest intensity 
when $\hbar\omega_i$ is tuned to the UHB pre-peak of the oxygen $K$-edge. 

In agreement with previous RIXS studies \cite{LearmonthRIXS,DudaRIXS} and 
ab initio quantum chemical calculations \cite{HozoiQC}, we assign the 
sharp peaks at about $\hbar\Omega=$2\,eV in \lico\  and 1.9\,eV in \cugeo\ 
to onsite $dd$-excitations, where the hole, which occupies the $3d_{x^2-y^2}$ 
orbital in the ground state, is excited to a different $3d$-level. 

In addition to this, well-resolved excitations are observed at resonance 
in between the $dd$-excitations and the fluorescence for both \lico\ 
($\hbar\Omega=3.2$\,eV) and \cugeo\ ($\hbar\Omega=3.8$\,eV), as 
indicated by arrows in Fig.\,\ref{fig_1}\,c and d. 
These two modes are essential for our further analysis. In the case 
of \cugeo, this excitation was observed previously and is known to be a 
ZRS exciton \cite{AtzkernEELS,BondinoRIXS, Kim2009}. Fig.\,\ref{fig_2} 
illustrates how such an exciton is created in the RIXS process at the oxygen $K$-edge.
Starting from two neighboring CuO$_4$ plaquettes $(d^9,d^9)$, the system 
reaches an intermediate state $(d^9,\underline{1s}d^{10})$ after absorbing 
the incoming photon tuned at the $1s\rightarrow 2p$ resonance of oxygen. 
In the final step, the $\underline{1s}$ oxygen core hole is filled by a 
ligand electron from the left plaquette, which results in a 
ZRS $d^9\underline{L}$ on this plaquette \cite{ZhangOrig} and a $d^{10}$ 
state on the right plaquette. The extra hole on the left plaquette and 
the extra electron  on  the right plaquette form a ZRS exciton. The total 
spin during this process is conserved at the oxygen $K$-edge.
Fig.\,\ref{fig_2} illustrates also that the RIXS intensity of this ZRS 
exciton will strongly depend on the orientation of the spins on 
neighboring CuO$_4$ plaquettes.

As we will show in the following, the 3.2\,eV excitation of \lico\ also 
corresponds to a ZRS exciton. In \lico, the situation is more controversial, 
because in previous experiments with RIXS \cite{LearmonthRIXS} and other 
experimental techniques \cite{AtzkernEELS2000,MizunoES,Kim2004}, the ZRS 
exciton could not be unambiguously observed.

\begin{figure}
\centering
\includegraphics[width=8.8cm]{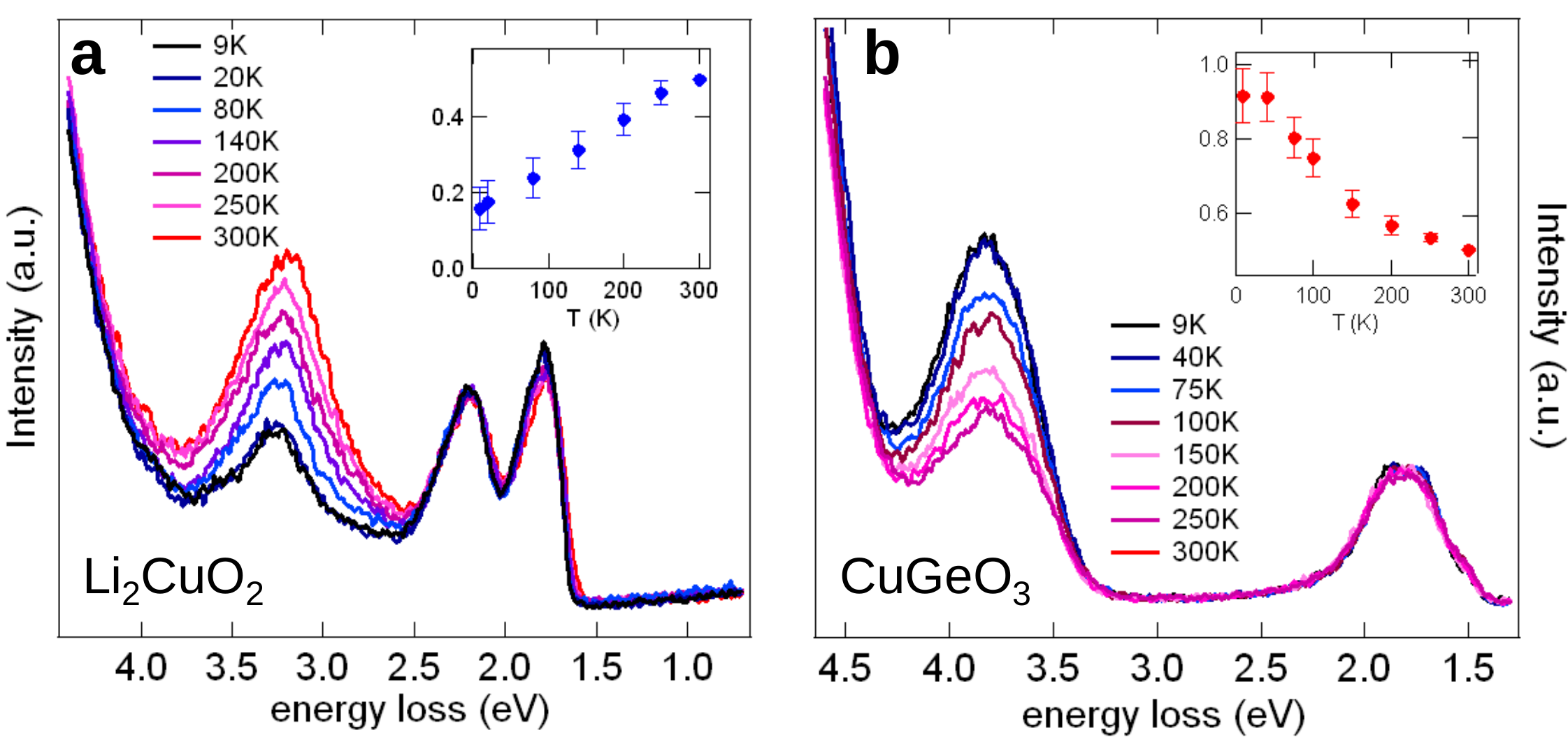}
\caption{\label{fig_3} 
(a) Temperature dependent RIXS data for \lico\ measured with $\sigma$-polarization 
and an excitation energy of 530.1 eV and (b) for \cugeo\ measured with 
$\pi-$polarization and an excitation energy of 530.8 eV. The spectra are 
plotted in an energy loss scale. The spectra have been normalised to the 
area of the $dd$-excitations (see Supplementary Material \cite{SMtheory}).}
\end{figure}

The most striking characteristic in the RIXS data is the dramatic $T$-dependence 
of the spectral peaks at about 3.5\,eV, which is present in both materials, 
see Fig. \ref{fig_3}.
Interestingly, the $T$-dependence in the two materials is opposite 
(Fig. \ref{fig_3} a and b):
in \LCO\ the exciton intensity decreases whereas in \CGO\ it increases upon cooling. 
These temperature dependences imply that high energy excitations at about 
3.5\,eV  are strongly affected by thermal fluctuations corresponding to an 
energy scale of merely $k_B\,T\sim1$\,meV. We will show that these 
high-energy modes thereby directly 
reflect the character of nearest neighbor spin correlations (see Fig.\ \ref{fig_2}), 
as the probability for a ZRS to be excited in RIXS strongly depends on the relative orientation of 
neighboring copper spins. In order to obtain a more detailed microscopic 
understanding of the nature of this strong temperature dependence, 
we performed many-body cluster calculations based on a $pd$-Hamiltonian 
for three up to five
\cuo\ plaquettes (trimers, tetramers, and pentamers)
\cite{Okada2007,OkadaZRSedge,OkadaZRScorner,MalekLCO} (see 
Ref.\ \onlinecite{SMtheory}
for more details). The use of a small cluster is 
justified by the fact that the electronic system of the edge-shared cuprates 
is well-localized. 

\begin{figure}
\centering
\includegraphics[width=9.0cm]{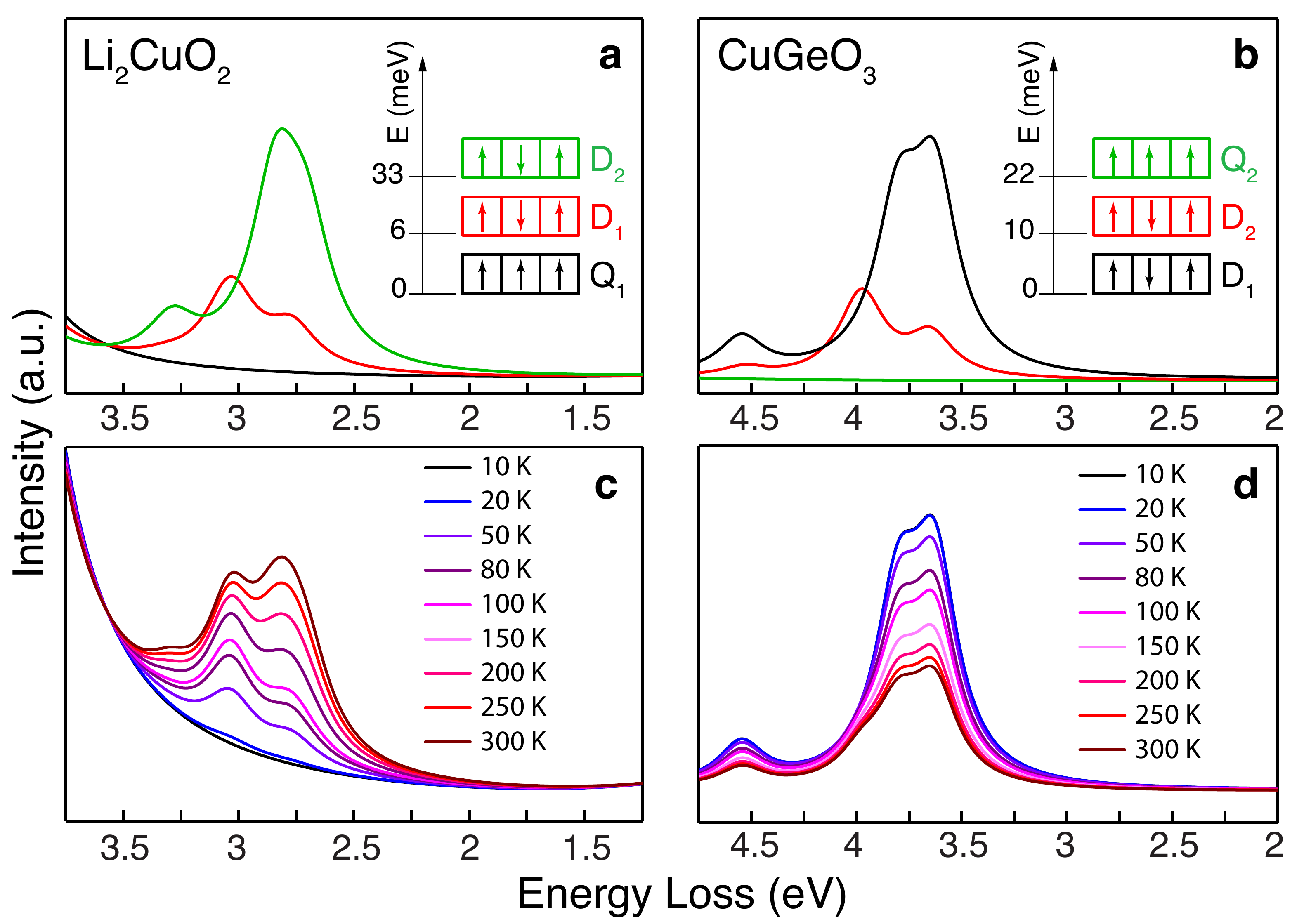}
\caption{\label{fig_4} 
(a) Partial RIXS intensity $\mathcal{I}_i$  calculated from the ground 
state (black line) of a Cu$_3$O$_{8}$ cluster and  the 
two excited states (red and green lines), respectively, for \lico\ and 
(b) for \cugeo. Using the same color code, a schematic representation of 
the eigenstates and corresponding eigenenergies is shown. (c), 
Total RIXS signal $\mathcal{I}(T)$ as a function of temperature 
for \lico\ and (d) \cugeo\ (Fig.\ S3 of the Supplementary material 
shows analogous results for a Cu$_5$O$_{12}$ cluster for comparison).
}
\end{figure}

We illustrate the underlying physics with the trimer
results for the sake of simplicity. Analogous results have been obtained on
tetramers (not shown in this work) and pentamers (as shown in Fig. S3 of the  
supplementary material \cite{SMtheory}).
The trimer  
has eigenstates $|i,S_i\rangle$ with total spin $S_i=1/2$ and 
$S_i=3/2$, corresponding to the spin configurations 
$\uparrow\downarrow\uparrow$ (AFM) and $\uparrow\uparrow\uparrow$ (FM), 
respectively. At finite temperature $T$, not only the ground state, but 
all eigenstates $|i,S_i\rangle$ within an energy range $\sim k_B\,T$ will 
be  populated. This includes states with different $S_i$ and corresponds 
to thermal spin fluctuations. For both \LCO\/ and \CGO\/  three $|i,S_i\rangle$ 
were found to be significantly populated within the studied temperature 
range (see Ref.\ \onlinecite{SMtheory} and Figs.\,\ref{fig_4} a,b): two doublets with $S_i=1/2$ ($D_{1,2})$, 
differing from each other in charge distribution among hybridised $p$ and 
$d$ states, and one quadruplet with $S_i=3/2$ ($Q_{1}$). However, their 
energy sequence is reversed for the two systems.

Each of the thermally populated $|i,S_i\rangle$ acts as an initial state 
for RIXS and opens specific excitation channels 
$|i,S_i\rangle \rightarrow |f,S_f\rangle$. Hence, every populated 
$|i,S_i\rangle $ contributes with a partial intensity $\mathcal{I}_i$, 
properly weighted in the total RIXS signal $\mathcal{I}(T)$ 
at a specific temperature $T$ as explained in Refs.\ \onlinecite{MalekLCO} 
and \onlinecite{SMtheory}. 

The calculated $\mathcal{I}_i$ for \LCO\/ and \CGO\/ are presented in 
Fig.\,\ref{fig_4}\,a and b, respectively. It can be seen that the 
$\mathcal{I}_i$ originating from $D_{1}$, $D_{2}$ and $Q_{1}$ are all 
distinctly different and, moreover, that a low energy charge transfer 
excitation exists, which can only be reached from $D_{1,2}$.
The calculations identify this excitation as the ZRS exciton  
$(d^9,d^9)\rightarrow (d^{10},d^9\underline{L})$ as illustrated in 
Fig.\,\ref{fig_2}.

The fact that the ZRS exciton can only be reached from the initial states 
$D_{1,2}$ and not from $Q_1$, is a direct consequence of the conservation 
of spin ($S_f=S_i$) in oxygen $K$-edge RIXS.
Such selection rules explain the strong 
$T$-dependence of the ZRS exciton 
intensity as shown in  Fig.\,\ref{fig_4}\,c and d, which is given by the thermal 
population of the excited multiplet 
$D$  states in case of \lico.

Both, the excitation energies and the $T$-dependence of the 
ZRS exciton obtained in the model calculations 
for \lico\ (Fig.\,\ref{fig_4}\,c) 
and \cugeo\ (Fig.\,\ref{fig_4}\,d), agree very well with the 
$T$-dependent peak found in our experiment 
(Fig.\,\ref{fig_3}\,a and b). This unambiguously identifies the 
experimentally observed excitations at 3.2\,eV in \LCO\ and 
at 3.8\,eV in \CGO\ as ZRS excitons.  

The fact that creating a ZRS exciton depends on the probability of two 
neighboring spins being antiparallel has an important consequence: 
it enables to study intrachain nearest neighbor spin correlations by 
oxygen $K$-edge RIXS. In \CGO, for instance, spins along the chain are antiparallel 
in the ground state. Upon decreasing $T$, 
thermally driven fluctuations out of this ground state therefore decrease, 
yielding a higher intensity of the corresponding ZRS exciton. Vice versa, 
if the spins along the chain are parallel in the ground state, the ZRS exciton 
peak becomes weaker upon cooling. 

The results in Fig.\,\ref{fig_3}\,a therefore not only resolve the issue 
of the ZRS exciton assignment in Li$_2$CuO$_2$, but also verify that the 
spins within chains in Li$_2$CuO$_2$ are FM ordered at low temperatures.
Another important observation for \lico\ is that even at 9\,K the ZRS 
exciton peak does not disappear, but retains a significant intensity 
(see Fig.\,\ref{fig_3} a) at variance with the weaker structure of the low-T spectrum 
calculated for pentamers (see our  Fig.\ S3 in Ref.\  \onlinecite{SMtheory}). This spectral structure is fully lacking for trimers as
shown in Fig.\,\ref{fig_4}\,c.

The observation of such a significant residual spectral weight of the 
ZRS exciton at low
temperatures is therefore surprising. The experimental results point to
sizeable residual 
 quantum and thermal fluctuations out of the intrachain FM ground state 
 and shows that excited spin flipped
  states must be very close in energy. 
More precisely, the presence of these excitations at 9\,K directly 
implies that the corresponding excitation energy must be less than 
$10$\,K$\simeq1$\,meV.
The present data therefore indicates
 that magnetic fluctuations in \lico\ 
persist down to low temperatures and that this system may be very close 
to the quantum 
critical point, which  separates FM from helical (AFM) intrachain order. 
This observation agrees very well with a previous neutron study, where 
the proximity to a quantum critical point was inferred from an analysis 
of the magnon 
dispersion\,\cite{LorenzINS}, whereas in the present case we observe the 
thermal and quantum nearest neighbor spin correlations directly.


Following Ref.\ \onlinecite{MalekLCO}, 
we remind that the 
Zeeman splitting of the non-singlet
excited states caused by an external magnetic field
affects thereby their thermal population. In the perspective of our present work, this suggests that RIXS measurements 
in magnetic fields
might be helpful
to resolve the nature of the involved spin states.

In conclusion, we have performed RIXS measurements at the oxygen $K$-edge 
on the edge-sharing chain compounds, \lico\ and \cugeo. Supported by 
 calculations within the five-band extended
Cu $3d$ O$2p$ Hubbard model, we have shown that our temperature dependent 
measurements give access to the nearest
neighbor spin correlations of these materials.
This is brought about via the entanglement of spin, orbital and charge 
degrees of freedom that characterizes strongly correlated, magnetically 
frustrated materials. RIXS at the oxygen $K$-edge can therefore be used 
as a versatile and powerful photon-in/photon-out method to investigate 
the low-energy magnetic
short-range spin fluctuations in large gap charge transfer insulators with 
great sensitivity.

V.B. and C.M., as well as J.G.\ and T.S., contributed 
equally to this work. 
This work was performed at the ADRESS beamline using the SAXES instrument 
jointly built by Paul Scherrer Institut, Switzerland and Politecnico di 
Milano, Italy.
This project was supported by the Swiss National Science Foundation and 
its National Centre of Competence in Research MaNEP as well as the Sinergia network MPBH. 
We acknowledge fruitful discussions with K.\ 
 Wohlfeld. V.B., R.K.\ and J.G.\  
gratefully acknowledge the financial support through the Emmy-Noether Program 
(Grant GE1647/2-1). V.B.\ also acknowledges the financial support from 
Deutscher Akademischer Austausch Dienst. S.J.\ acknowledges financial 
support from the Foundation for Fundamental Research on Matter 
(The Netherlands).

\end{document}